# Molecular Gas in the Inner 3.2 Kpc of NGC 2403: Star Formation at Subcritical Gas Surface Densities

Michele D. Thornley[1]
Department of Astronomy, University of Maryland
College Park, MD 20742-2421

and

Christine D. Wilson[2]
Department of Physics and Astronomy, McMaster University
Hamilton, Ontario, Canada L8S 4M1

## ABSTRACT

We present a fully sampled map of the inner 3.2 kpc of the nearby spiral galaxy NGC 2403 in the CO J=1-0 line. These data emphasize the relatively small contribution of molecular hydrogen to the cold gas content of this galaxy: in the inner kiloparsec the molecular hydrogen surface density is 0.6 times that of atomic hydrogen, in contrast with most spiral galaxies for which molecular hydrogen is the dominant phase in the inner regions. These data confirm that the gas surface densities in the inner 2.8 kpc of NGC 2403 lie below the critical surface density for star formation under the theory proposed by Kennicutt (1989). Since star formation is occurring throughout the inner disk, despite the subcritical gas surface densities, this simple dynamical model cannot be the only important process regulating star formation in galaxies. This conclusion is supported by the widespread star formation seen at subcritical gas densities in the inner disks of M33 and some Sa galaxies. We suggest that stochastic star formation processes are responsible for the star formation seen in these regions and thus that supercritical gas densities may not be a necessary condition for star formation in the inner regions of galactic disks.

*Subject headings:* galaxies: individual(NGC 2403) — galaxies: spiral — galaxies: interstellar matter — ISM: molecules

[1] michele@astro.umd.edu

[2] wilson@eccles.physics.mcmaster.ca

## 1. Introduction

In recent years, there have been several attempts to formulate a universal star formation law for galactic disks. Studies have revealed correlations of star formation activity with metallicity (Rana & Wilkinson 1986), local angular frequency ( Wyse & Silk 1989), and gas surface density (e.g., Wyse & Silk 1989, Kennicutt 1989). Kennicutt (1989) proposed a non-linear star formation law for spiral galaxies which depends on a surface density threshold, with the stability criteria for a thin gas disk as the physical basis for this threshold. Kennicutt found that the familiar modified Schmidt power law relationship between the radial distributions of star formation and cold gas surface density breaks down for gas surface densities below a threshold value in his sample of 15 spiral galaxies. The Kennicutt model is very successful in providing a physical basis for the lower star formation activity in early-type galaxies, the non-linear increase in star formation in the spiral arms of some galaxies, and the outer radial cutoff in star forming disks.

Two galaxies that may be exceptions to Kennicutt's star formation law are M33 and NGC 2403, which appear to have subcritical gas surface densities over a significant fraction of their star forming disks. As both M33 and NGC 2403 have low surface brightness and are relatively slow rotators, these results suggest that Kennicutt's star formation law may not apply to all galaxy types. However, the CO data used in Kennicutt's analysis consisted of coarsely sampled observations along the major axis (Young 1987), and so the molecular surface density may have been underestimated. A fully sampled map of M33 increased the estimate of total molecular gas content (Wilson & Scoville 1989), although the increase was insufficient to raise the gas surface density above the critical level (Wilson, Scoville & Rice 1991).

In this paper, we present a fully sampled map of $^{12}$CO J=1-0 emission from the inner 3.2 kpc of NGC 2403. This more complete assessment of the cold (atomic + molecular) gas component of NGC 2403 is used to test the star formation law proposed by Kennicutt (1989). The observations and data reduction are discussed in §2. The large scale properties of the molecular gas in this region are discussed and compared with previous estimates in §3. The comparison of Kennicutt's threshold surface density profile with the observed cold gas surface density distribution in the inner 3.2 kpc of NGC 2403 is discussed in §4. The paper is summarized in §5.

## 2. Observations and Data Reduction

A fully sampled map of the inner 3.2 kpc of NGC 2403 was obtained at the NRAO 12 meter telescope[3] in observing runs in 1991 October, 1993 May, and 1994 October. At the assumed distance of 3.2 Mpc (Freedman & Madore 1988), the beam of the 12m telescope (FWHM=54″) subtends a linear scale of 840 pc. A total of 81 positions at 30″ spacing were observed, and five to eight six-minute scans were taken for each position. The 256 channel 1 MHz (2.6 km s$^{-1}$) filterbank was configured in parallel mode to detect both polarizations. Pointing curves showed small slopes in both azimuth and elevation, with peak to peak scatter of ±10″. Observations of IRC+10216 and Orion A were used to check for variations in the calibration of the temperature scale between the three observing runs. Peak strengths of these sources observed in 1993 and 1994 were within 20% of the peak strengths observed in 1991.

The data were fit with linear baselines and the two polarizations were averaged together. The averaged scans for each position were coadded and binned to 5.2 km s$^{-1}$ velocity resolution. The temperatures were converted to a main-beam temperature scale ($T_{mb}=T_R^*/\eta_m^*$, where $\eta_m^*$=0.82 at 115 GHz) and the rms noise ($\sigma$) was measured for each output spectrum. Integrated intensities and line widths were determined for all positions where both the peak line temperature and the integrated intensity were greater than three times the uncertainty ($3\sigma$ and $3\sigma_{line}$, respectively). Spectra which did not match these criteria were counted as non-detections, and upper limits to the integrated intensity were measured by integrating over a similar velocity range as that used for nearby detections. The uncertainty in the integrated intensity is given by $\sigma_{line} = \sigma(\Delta V)\sqrt{N_{line}(1 + \frac{N_{line}}{N_{base}})}$, where $\sigma$ is the rms noise, $\Delta V$ is the velocity width per channel, $N_{line}$ is the number of channels over which the line was integrated, and $N_{base}$ is the number of channels used to determine the linear baseline. The rms noise, peak temperature, centroid velocity, velocity width, and integrated intensity for each position are given in Table 1, and the observed spectra are shown in Figure 1.

## 3. Large Scale Properties of the CO Emission

The distribution of emission in the inner 3.2 kpc of NGC 2403 is shown in Figure 1. Weak emission is detected over much of the map, and there are three regions of particular interest. The strongest peak lies northwest of the nucleus, along the major axis. This emission is likely associated with the third brightest HII region in NGC 2403 (Kennicutt 1988). Although most of the map shows an exponential decrease of emission with increasing radius, the spectra on the northeast edge of this map are relatively strong. The brightest HII region in NGC 2403 lies just inside the northeast edge of the map. Lastly, there are relatively bright peaks near the northwestern end of the major axis, likely associated with the second brightest HII region, which is just off the edge of the existing map. The properties of these positions, together with those of other strong lines, will be discussed in more detail in §4.

---

[3]Operated by Associated Universities, Inc., under cooperative agreement with the National Science Foundation.

To determine the CO radial profile, the data were azimuthally averaged in elliptical annuli spaced by 30″ along the minor axis of the galaxy (PA 125°, $i$=60°, Shostak 1973). The column density in the plane of the galaxy was derived from the average integrated intensity in each annulus using the expression $N_{H_2} = 3 \times 10^{20} \cos(i) I_{CO}$, where $3 \times 10^{20}$ cm$^{-2}$ (K km s$^{-1}$)$^{-1}$ is the CO-to-H$_2$ conversion factor (Strong et al. 1988; Scoville & Sanders 1987) and $i$ is the inclination of the galaxy. The calculated surface densities for each annulus are shown in Table 2, and our radial profile is compared with that of Young (1987) in Figure 2. The profile derived from the fully sampled map is somewhat smoother than that derived from the major axis observations, with a central value 1.2 times lower than that in the Young profile and a scale length twice as long (2.5 kpc). Based on the data published by Young, it is not clear what caused the discrepancy between the two profiles; it may be partly due to the difference in beam sizes (54″ vs. 45″ FWHM), resulting in an increase in the flux of the central position in Young's data, and partly due to the undersampled nature of the earlier map. However, the total mass in the inner kiloparsec calculated from each dataset is similar, both yielding approximately $1.5 \times 10^7$ M$_\odot$.

Using the new CO radial profile and the HI profile of Wevers, van der Kruit, & Allen (1986), we have examined the molecular, atomic, and total cold gas content of the inner kiloparsec of NGC 2403. These data emphasize the difference between the cold gas properties of low mass galaxies like NGC 2403 and M33 and those of more massive spirals. From the rotation curve from Shostak (1973), the total mass of NGC 2403 inside one kiloparsec is $7.5 \times 10^8$ M$_\odot$. The masses of H$_2$ and HI in this region are $1.5 \times 10^7$ M$_\odot$ and $2.5 \times 10^7$ M$_\odot$ (Wevers et al 1986), respectively, and thus the cold gas makes up 5% of the total mass. The cold gas fraction and the total mass of the inner kiloparsec of M33 are similar to those of NGC 2403 (Wilson & Scoville 1989; see correction to H$_2$ masses in Thornley & Wilson 1994). Further, the value of the molecular to atomic mass ratio (H$_2$/HI) in both galaxies is consistent with unity: 1.3 in M33 and 0.6 in NGC 2403. Though this may indicate that the gas in the inner disk of M33 is mostly molecular while the gas in the inner disk of NGC 2403 is mostly atomic, the H$_2$/HI ratio in both galaxies is more than an order of magnitude less than the ratios seen in the inner disks of more massive spirals (e.g., M51, Lord & Young 1990; NGC 6946, Tacconi & Young 1989). Such low ratios are likely an effect of the low gas surface densities in NGC 2403 and M33. If all galaxies require a similar shielding column density of HI to form $H_2$, then galaxies with low surface densities will have a smaller fraction of their interstellar medium in the form of molecular gas than galaxies of higher surface density (Wilson and Scoville 1989).

## 4. Star Formation and the Surface Density Threshold in NGC 2403

### 4.1. Testing the Threshold Value with Azimuthal Averages

Although molecular hydrogen constitutes a small fraction of the total cold gas in NGC 2403, it is important in the inner disk, where it fills in some of the HI depression in the inner ~3.2 kpc (see Figure 3). The more complete molecular and atomic data set is used here to reexamine Kennicutt's analysis of the star formation conditions in NGC 2403. To be consistent with Kennicutt's derivation, we have calculated the total gas surface density by multiplying the total hydrogen surface density (atomic + molecular) by 1.45 to include the presence of other elements. The measured molecular and atomic hydrogen surface densities and the resulting total gas surface density at each radius are shown in Table 2.

Kennicutt's (1989, 1990) formulation of the critical density is given by

$$\Sigma_c = \alpha \frac{\kappa c}{\pi G} \qquad (1)$$

where $\kappa = 1.41 \frac{V}{R}\sqrt{(1+\frac{R}{V}\frac{dV}{dR})}$ is the epicyclic frequency and $c$ is the velocity dispersion of the gas. The dimensionless quantity $\alpha$ is of order unity, and describes the properties of the disk. Kennicutt compared the ratio of gas to critical density to the radius of the disk of HII regions to empirically determine a value of $\alpha$. For comparison with Kennicutt's original results, we will use $\alpha=0.63$ and $c=6$ km s$^{-1}$[4]. We have used the parametrization of the rotation curve of NGC 2403 derived by Shostak (1973),

$$V(r) = \frac{v_t(r/r_t)}{[\frac{1}{3} + \frac{2}{3}(r/r_t)^n]^{\frac{3}{2n}}} \qquad (2)$$

where $v_t = 126$ km s$^{-1}$, $r_t = 9'$, and $n = 1.0$, to calculate the value of $\kappa$ at each radius. This differs slightly from Kennicutt's method, which used a fit to the rotation curve consisting of a linearly rising inner portion and a flat outer portion. Table 3 shows the epicyclic frequency and critical surface density at each radius, as well as the total gas surface density from Table 2. A comparison of the gas and critical densities in Table 3 shows that the inner regions of NGC 2403 have subcritical surface densities, while in the last annulus the surface density exceeds the critical surface density by more than 10%. Extrapolating the distribution of CO from an exponential fit to our radial profile indicates that the gas surface density should exceed the critical density at a radius of 2.8 kpc, approximately the same radius as the brightest HII region in NGC 2403 (Kennicutt 1988). The HI alone exceeds the critical surface density at 3.4 kpc, similar to the radius at which the total cold gas component becomes supercritical in the study by Kennicutt (1989, Fig. 11). Figure 3 compares the radial profiles of the atomic, total (atomic+molecular), and critical surface densities. Thus, using more complete CO data, we confirm that NGC 2403 has surface densities below the critical value in the inner 2.8 kpc, despite the vigorous star formation occurring there.

---

[4]The value of $\alpha$ used here has been adjusted from the published value of $\alpha=0.67$ (Kennicutt 1989), which was empirically determined from a slightly different critical density expression based on the stability criterion for a stellar disk (see Kennicutt 1989,1990).

Can we increase the surface densities to supercritical values in the inner disk by changing any of the assumptions made in this analysis? As a way to bring M33 and NGC 2403 in line with the rest of his sample, Kennicutt suggested that the CO-to-$H_2$ conversion factor might differ from the Galactic value in these galaxies. The CO-to-$H_2$ conversion factor would have to be larger by a factor of four before the observed gas surface densities would exceed the critical density in the inner 470 pc. However, a comparison of the virial and molecular masses of nine well-resolved molecular clouds in M33 indicates that the value of the CO-to-$H_2$ conversion factor in the inner disk of M33 is similar to that in our own Galaxy (Wilson & Scoville 1990). As the inner disk of NGC 2403 has a similar metallicity to the inner disk of M33 (Tosi & Diaz 1985), there is no evidence to support a significantly higher conversion factor based on metallicity differences. Further, the CO-to-$H_2$ conversion factor in the SMC, where the metallicity is one-tenth of the solar value, is just four times the Galactic value (Rubio, Lequeux, & Boulanger 1993). It is also possible that the surface density of atomic gas is underestimated by the usual assumption of optically thin gas in the conversion of brightness temperature to surface density. A recent study of HI emission and absorption in the Milky Way and M31 (Braun & Walterbos 1992) suggests that the amount of atomic gas in the Galaxy may be underestimated by as much as 30% in the optically thin assumption for emission regions with a brightness temperature of 50 K. This factor would not significantly change the radius at which the gas surface density exceeds the critical surface density in NGC 2403. For brightness temperatures of 100 K, HI surface densities in the Milky Way may be underestimated by as much as 130% (Braun & Walterbos 1992). The underestimates in M31 are approximately a factor of two lower than those quoted for the Galaxy. If we estimate that the average HI surface density in NGC 2403 is underestimated by as much as a factor of two, the radius at which $\Sigma_g = \Sigma_c$ decreases to a kiloparsec. Thus, the estimate of higher atomic gas surface density of optically thick gas is not sufficient to raise the total gas surface density above the critical surface density in the inner kiloparsec.

Another possibility is the velocity dispersion, $c$. In a more slowly rotating galaxy such as NGC 2403 ($v_{max}=v_t=$ 126 km s$^{-1}$, Shostak 1973), it is possible that the velocity dispersion is lower than in the larger galaxies of Kennicutt's sample. Assuming a Galactic CO-to-$H_2$ conversion factor, the velocity dispersion in NGC 2403 would have to be 3 km s$^{-1}$ for the critical surface density to be brought down to the observed gas surface density in the center of NGC 2403. This value is at the lower end of the range of measured velocity dispersions cited by Kennicutt (3-10 km s$^{-1}$). The cloud-to-cloud velocity dispersion of giant molecular clouds in M33 is 5±1 km s$^{-1}$ (Wilson & Scoville 1990), in good agreement with values measured for giant molecular clouds in the more rapidly rotating Milky Way. Given that the maximum rotation velocity in M33 is even lower than that of NGC 2403 (103 km s$^{-1}$, Rogstad & Shostak 1972), there is no compelling evidence that the velocity dispersion in NGC 2403 should be significantly different from the assumed value of 6 km s$^{-1}$. In addition, the value of $\alpha$ determined by fitting the outer cutoff to the HII region disk with $c = 6$ km s$^{-1}$ appears appropriate for NGC 2403 (Kennicutt 1989, Fig. 11). Thus any decrease in $c$ would require a corresponding increase in $\alpha$ in order to fit the HII disk cutoff, and so would produce no change in the calculated critical densities.

### 4.2. Testing the threshold value on smaller scales

Although the radially averaged gas surface density is subcritical in the inner disk of NGC 2403, there may be individual regions which are supercritical on scales of the beam size. By testing individual regions, we can see if Kennicutt's star formation law holds on kiloparsec scales as well as whether star formation in the inner disk occurs in localized regions of supercritical surface density. Using a recently observed HI map (R. Braun 1994, private communication), we have calculated the average HI surface density within a 54″ (840 pc) beam at each of the 15 positions with the strongest CO emission. The positions and their observed and critical surface densities are shown in Table 4. At eight positions the observed gas surface density is within 20% of or exceeds the critical surface density at that radius; given the calibration uncertainties, these values are consistent with supercritical densities. Four of the eight are at radii greater than 2 kpc, and two of these are likely associated with two of the three brightest HII regions (Kennicutt 1988). As the radial profile indicates that the average total gas surface density is supercritical at radii larger than 2.8 kpc, it is likely that most positions on the edge of the map represent supercritical regions.

A comparison of the CO and Hα profiles also suggests adherence to a "normal" star formation law farther out in the disk. Figure 4 compares the Hα profile from Kennicutt (1989) with the observed CO profile and an exponential fit to the outer points of our radial CO profile. This exponential CO profile reproduces the observed value of the surface density at the center of the galaxy to within a few percent, and allows us to extrapolate the CO disk to larger radii. An exponential fit to the inner 4 kpc of the Hα radial profile gives an Hα scale length of approximately 2.5 kpc, similar to that of the CO radial profile presented in this paper. Kennicutt (1989) noted that galaxies like M51 and NGC 6946 have molecular surface density profiles which closely follow the surface brightness profile of Hα emission, while in CO-weak galaxies like NGC 2403, the two profiles appear to be relatively uncorrelated. The correlation between the CO and Hα profiles presented here increases with radius, which may indicate that star formation at greater radii is governed by a more standard star formation law.

### 4.3. Star formation at dynamically subcritical densities

The question of how stars form in the subcritical inner disk must still be addressed. The remaining four positions of supercritical surface density are located between 1.1 and 1.6 kpc from the center position. These positions may represent localized regions of supercritical gas density; indeed, the third brightest HII region in NGC 2403 is likely associated with the innermost of the supercritical positions. However, the larger number of subcritical points suggests that the gas in the inner disk is not sufficiently dense, even on kiloparsec scales, for star formation to occur. The observation that star formation is common in the inner disk of NGC 2403, despite a gas surface density that is subcritical over most of this area, suggests that the simple dynamical theory proposed by Kennicutt (1989) is not the only important process involved in regulating star formation in galaxies. Additional evidence comes from the star formation occurring at subcritical gas densities in the inner disk of M33 (Wilson et al. 1991) and the large numbers of HII regions identified using sensitive Fabry-Perot observations of the inner disks of Sa galaxies, where the gas is also subcritical (Caldwell et al. 1991). Although Kennicutt's dynamical model is very successful in explaining the outer radial cutoff of star formation in disk galaxies, it is less successful in predicting where star formation can occur in the inner parts of disk galaxies, particularly when they have a low total mass or are of early type.

Stochastic star formation processes are also likely to be important in galaxies (see review by Elmegreen 1995). An example of such a model is that of Chiang & Prendergast (1985), which models a two-fluid star-gas system that is initially in an unstable equilibrium, with star formation balancing stellar mass loss and gas cooling balancing stellar heating. This simple model assumes a star formation rate that is proportional to the gas volume density and produces traveling waves of star formation which arise from excess heating and compression of the gas by newly formed stars. However, this model is not entirely realistic as it neglects the molecular phase of the interstellar medium and does not include self-gravity between the fluid elements. Such stochastic processes are likely to be most important in regions of galactic disks with relatively high gas and stellar surface densities, where the propagating or destabilizing effects of star formation can be transmitted most efficiently from one region to another. If both dynamical and stochastic star formation processes are important in galactic disks, dynamical processes are likely to dominate in the outer regions, where the surface density is too low for efficient destabilization by star formation. Thus in the outer disks, star formation would only occur where the gas is dynamically unstable to the growth of large-scale density perturbations (as in Kennicutt 1989). However, in the inner disk, local processes can become important because of higher surface densities of both stars and gas, and thus some star formation can occur even if the gas is below the dynamically-determined critical density. It may be that only relatively low star formation rates can be supported by stochastic star formation: the total star formation rate for NGC 2403 is only 0.1 $M_\odot$ yr$^{-1}$ (estimated from Fig. 2 of Kennicutt 1989). Stochastic processes may also be the dominant processes in regulating star formation in dwarf and irregular galaxies, which have much less rotational shear and lower masses than large spiral galaxies.

## 5. Conclusions

A fully sampled map of the inner 3.2 kpc of the nearby spiral galaxy NGC 2403 in the J=1-0 line emphasizes the relatively small contribution of molecular hydrogen to the cold gas content of this galaxy. In the inner kiloparsec, the surface density of molecular hydrogen is 0.6 times that of atomic hydrogen, in sharp contrast to other spiral galaxies where molecular hydrogen dominates in the inner regions. A similar $H_2$/HI ratio is observed in the inner disk of M33, suggesting that the formation of molecular hydrogen in these low mass galaxies is constrained by the low surface density of the gas, which provides a limited amount of HI to shield the formation of $H_2$.

Combining these new CO data with published HI data confirms that the gas surface densities in the inner disk are below the threshold surface density for star formation, according to the star formation law proposed by Kennicutt (1989). The radial profiles suggest that the total gas surface density becomes supercritical at R=2.8 kpc, while the HI surface density alone is supercritical at R=3.4 kpc. To make the observed gas surface density and the threshold surface density comparable in the inner 0.5 kpc of the galaxy would require a CO-to-$H_2$ conversion factor four times larger than the Galactic value, or a velocity dispersion a factor of two lower than that assumed by Kennicutt. Comparison of NGC 2403 with M33, a galaxy with similar metallicity and inner rotation curve for which both the conversion factor and the velocity dispersion have been measured, suggests that both these options are unlikely. Further, assuming that the HI surface density is underestimated by as much as a factor of two due to the assumption of optically thin gas in the calculation of HI surface densities is not sufficient to raise gas surface densities above the critical surface density in the central kiloparsec.

Eight individual regions were found to have marginally critical gas surface densities, when the gas surface density was measured over the scale of the CO beam (54″=840 pc). Four of these positions are at radii greater than 2 kpc. It is likely that the gas at all positions at the edge of the map are in regions with conditions which obey a star formation law such as that formulated by Kennicutt; the agreement of CO and H$\alpha$ radial profiles at larger radii supports this suggestion. However, few positions in the inner disk are indicative of supercritical surface densities. Therefore, the presence of star formation throughout the inner disk of NGC 2403 implies that the simple dynamical model proposed by Kennicutt (1989) is not the only important process involved in regulating star formation in galaxies. This result is supported by the star formation at subcritical densities seen in the inner disk of M33 (Wilson et al. 1991) and the inner disks of Sa galaxies (Caldwell et al. 1991). Star formation in these regions is likely due to stochastic star formation processes, which may be most important in the inner disks of galaxies where the gas and stellar surface densities are higher than in the outer disks.

Many thanks to Robert Braun, for allowing use of his HI mosaic of NGC 2403 prior to publication. We would also like to thank Robert Kennicutt for helpful correspondence about his formulation for the star formation threshold surface density. Partial support for this research was provided by NSF grant AST-9314847 and the State of Maryland through its contributions to the Laboratory for Millimeter-Wave Astronomy (M.D.T.), and by NSERC (Canada) through a Women's Faculty Award and Research Grant (C.D.W.).

– 11 –Table 1. $^{12}$CO Emission in NGC 2403

| $\Delta\alpha^a$ (′) | $\Delta\delta^a$ (′) | $T_{peak}^b$ (mK) | $V_{cent}^c$ (km s$^{-1}$) | $\Delta V_{FWHM}^d$ (km s$^{-1}$) | $\int T_{mb}\,dV^e$ (K km s$^{-1}$) |
|---|---|---|---|---|---|
| 2.5 | 0.0 | 26± 9 | ⋯ | ⋯ | (0.94±0.22) |
| 2.5 | -0.5 | 59± 13 | 209 | 23 | 1.30±0.22 |
| 2.5 | -1.0 | 34± 12 | ⋯ | ⋯ | (0.81±0.20) |
| 2.5 | -1.5 | 24± 15 | ⋯ | ⋯ | (-0.10±0.30) |
| 2.5 | -2.0 | 30± 13 | ⋯ | ⋯ | (0.21±0.24) |
| 2.0 | 0.5 | 51± 14 | 175 | 30 | 1.10±0.21 |
| 2.0 | 0.0 | 62± 14 | 193 | 26 | 1.66±0.27 |
| 2.0 | -0.5 | 82± 15 | 202 | 21 | 1.53±0.23 |
| 2.0 | -1.0 | 52± 16 | 211 | 18 | 0.99±0.24 |
| 2.0 | -1.5 | 56± 20 | ⋯ | ⋯ | (1.22±0.36) |
| 2.0 | -2.0 | 44± 12 | 211 | 29 | 1.14±0.24 |
| 1.5 | 1.0 | 39± 17 | ⋯ | ⋯ | (0.71±0.45) |
| 1.5 | 0.5 | 67± 10 | 168 | 26 | 1.38±0.17 |
| 1.5 | 0.0 | 58± 14 | 183 | 52 | 2.12±0.30 |
| 1.5 | -0.5 | 65± 15 | 199 | 42 | 1.80±0.30 |
| 1.5 | -1.0 | 82± 19 | 210 | 21 | 1.66±0.29 |
| 1.5 | -1.5 | 65± 18 | 209 | 19 | 1.34±0.27 |
| 1.5 | -2.0 | 44± 11 | 197 | 30 | 0.99±0.19 |
| 1.0 | 1.5 | 26± 14 | ⋯ | ⋯ | (0.55±0.39) |
| 1.0 | 1.0 | 74± 21 | 130 | 36 | 2.22±0.43 |
| 1.0 | 0.5 | 55± 18 | 155 | 35 | 1.63±0.39 |
| 1.0 | 0.0 | 74± 17 | 173 | 41 | 2.48±0.36 |
| 1.0 | -0.5 | 71± 13 | 202 | 26 | 2.29±0.28 |
| 1.0 | -1.0 | 56± 8 | 209 | 24 | 1.32±0.13 |
| 1.0 | -1.5 | 51± 22 | ⋯ | ⋯ | (0.68±0.35) |
| 1.0 | -2.0 | 33± 14 | ⋯ | ⋯ | (0.52±0.27) |
| 0.5 | 1.5 | 36± 9 | 121 | 28 | 0.83±0.15 |
| 0.5 | 1.0 | 50± 13 | 122 | 23 | 1.16±0.20 |
| 0.5 | 0.5 | 90± 22 | 158 | 30 | 1.99±0.46 |
| 0.5 | 0.0 | 74± 17 | 167 | 56 | 1.98±0.35 |
| 0.5 | -0.5 | 85± 17 | 183 | 39 | 2.87±0.41 |
| 0.5 | -1.0 | 55± 13 | 193 | 32 | 1.92±0.34 |
| 0.5 | -1.5 | 17± 17 | ⋯ | ⋯ | (-0.27±0.38) |
| 0.5 | -2.0 | 32± 10 | 187 | 34 | 0.79±0.18 |

Table 1—Continued

| $\Delta\alpha$[a] (′) | $\Delta\delta$[a] (′) | $T_{peak}$[b] (mK) | $V_{cent}$[c] (km s$^{-1}$) | $\Delta V_{FWHM}$[d] (km s$^{-1}$) | $\int T_{mb}$ dV[e] (K km s$^{-1}$) |
|---|---|---|---|---|---|
| .0 | 2.0 | 18± 10 | ⋯ | ⋯ | (0.49±0.29) |
| 0.0 | 1.5 | 49± 15 | ⋯ | ⋯ | (0.60±0.27) |
| 0.0 | 1.0 | 52± 12 | 106 | 53 | 0.96±0.24 |
| 0.0 | 0.5 | 44± 18 | ⋯ | ⋯ | (1.15±0.55) |
| 0.0 | 0.0 | 41± 9 | 136 | 66 | 2.55±0.27 |
| 0.0 | -0.5 | 60± 15 | 159 | 80 | 2.81±0.41 |
| 0.0 | -1.0 | 82± 10 | 179 | 61 | 2.61±0.23 |
| 0.0 | -1.5 | 38± 13 | ⋯ | ⋯ | (0.79±0.21) |
| 0.0 | -2.0 | 39± 9 | 149 | 32 | 1.08±0.18 |
| -0.5 | 2.0 | 39± 13 | ⋯ | ⋯ | (0.25±0.24) |
| -0.5 | 1.5 | 35± 17 | ⋯ | ⋯ | (0.28±0.46) |
| -0.5 | 1.0 | 59± 18 | 86 | 44 | 1.77±0.58 |
| -0.5 | 0.5 | 82± 23 | 86 | 37 | 2.65±0.55 |
| -0.5 | 0.0 | 75± 13 | 95 | 63 | 2.63±0.33 |
| -0.5 | -0.5 | 77± 12 | 131 | 29 | 1.75±0.24 |
| -0.5 | -1.0 | 55± 14 | 141 | 45 | 1.88±0.30 |
| -0.5 | -1.5 | 44± 12 | 135 | 23 | 0.86±0.18 |
| -1.0 | 2.0 | 31± 13 | ⋯ | ⋯ | (0.80±0.22) |
| -1.0 | 1.5 | 37± 16 | ⋯ | ⋯ | (1.19±0.29) |
| -1.0 | 1.0 | 71± 15 | 67 | 20 | 1.50±0.27 |
| -1.0 | 0.5 | 133± 18 | 72 | 28 | 3.62±0.36 |
| -1.0 | 0.0 | 62± 16 | ⋯ | ⋯ | (1.06±0.36) |
| -1.0 | -0.5 | 33± 11 | 105 | 72 | 1.60±0.37 |
| -1.0 | -1.0 | 46± 16 | ⋯ | ⋯ | (0.83±0.36) |
| -1.0 | -1.5 | 47± 10 | 129 | 18 | 0.87±0.22 |
| -1.5 | 2.0 | 37± 10 | 74 | 31 | 0.85±0.21 |
| -1.5 | 1.5 | 54± 17 | 64 | 41 | 1.32±0.33 |
| -1.5 | 1.0 | 69± 13 | 62 | 22 | 1.49±0.20 |
| -1.5 | 0.5 | 57± 11 | 68 | 27 | 1.56±0.19 |
| -1.5 | 0.0 | 65± 25 | ⋯ | ⋯ | (0.95±0.41) |
| -1.5 | -0.5 | 38± 15 | ⋯ | ⋯ | (1.13±0.34) |
| -1.5 | -1.0 | 42± 13 | 103 | 23 | 0.94±0.23 |
| -2.0 | 2.0 | 40± 10 | 50 | 18 | 0.68±0.19 |
| -2.0 | 1.5 | 49± 21 | ⋯ | ⋯ | (-0.16±0.57) |

Table 1—Continued

| $\Delta\alpha^a$ (′) | $\Delta\delta^a$ (′) | $T_{peak}^b$ (mK) | $V_{cent}^c$ (km s$^{-1}$) | $\Delta V_{FWHM}^d$ (km s$^{-1}$) | $\int T_{mb}\,dV^e$ (K km s$^{-1}$) |
|---|---|---|---|---|---|
| -2.0 | 1.0  | 43± 16 | ⋯ | ⋯ | (0.73±0.35) |
| -2.0 | 0.5  | 15± 17 | ⋯ | ⋯ | (-0.56±0.37) |
| -2.0 | 0.0  | 39± 22 | ⋯ | ⋯ | (-0.35±0.49) |
| -2.0 | -0.5 | 51± 18 | ⋯ | ⋯ | (1.09±0.29) |
| -2.5 | 2.0  | 37± 13 | ⋯ | ⋯ | (1.12±0.25) |
| -2.5 | 1.5  | 64± 12 | 45 | 14 | 0.86±0.21 |
| -2.5 | 1.0  | 52± 10 | 58 | 31 | 1.62±0.20 |
| -2.5 | 0.5  | 64± 13 | 64 | 26 | 1.53±0.28 |
| -2.5 | 0.0  | 54± 12 | 65 | 13 | 0.81±0.16 |
| -3.0 | 1.5  | 78± 10 | 44 | 23 | 1.76±0.19 |
| -3.0 | 1.0  | 55± 12 | 55 | 28 | 1.48±0.24 |
| -3.0 | 0.5  | 45± 12 | 58 | 20 | 0.89±0.19 |
| -3.0 | 0.0  | 42± 14 | 58 | 20 | 0.70±0.21 |

[a] Cols. (1,2) – Right ascension and declination offsets are relative to $07^h\ 32^m\ 01.2^s$, $65°\ 42'45''$ (1950).

[b] Col. (3) – Uncertainty in peak temperature is $1\sigma$. Temperatures are on the main beam brightness scale.

[c] Col. (4) – Centroid velocity for range of integration.

[d] Col. (5) – FWHM of gaussian fit to line profile.

[e] Col. (6) – Values in parentheses are upper limits. See §2 for calculation of uncertainties and upper limits.

Table 2: Molecular and Atomic Surface Densities

| R(kpc)[a] | $2N_{H_2}$ ($10^{20}$ H atoms cm$^{-2}$) | $\Sigma_{H_2}$ (M$_\odot$ pc$^{-2}$) | $\Sigma_{HI}$ (M$_\odot$ pc$^{-2}$) | $\Sigma_g$[b](M$_\odot$ pc$^{-2}$) |
|---|---|---|---|---|
| 0 | 7.65 | 6.12 | 8.22 | 20.79 |
| 0.47 | 6.35 | 5.09 | 8.18 | 19.24 |
| 1.40 | 4.92 | 3.94 | 8.09 | 17.44 |
| 2.33 | 2.79 | 2.23 | 8.25 | 15.20 |
| 3.26 | 2.25 | 1.80 | 9.24 | 16.01 |

[a] Col. (1) – d = 3.2 Mpc.
[b] Col. (5) – $\Sigma_g = 1.45\,(\Sigma_{H_2} + \Sigma_{HI})$.

Table 3: Comparison of Critical and Observed Gas Surface Densities

| R (kpc)[a] | $\kappa$ (km s$^{-1}$kpc$^{-1}$)[b] | $\Sigma_c$ (M$_\odot$ pc$^{-2}$)[c] | $\Sigma_g$ (M$_\odot$ pc$^{-2}$)[d] |
|---|---|---|---|
| 0 | 145.5 | 40.4 | 20.8 |
| 0.47 | 127.6 | 33.2 | 19.2 |
| 1.40 | 91.2 | 23.7 | 17.4 |
| 2.33 | 68.7 | 17.8 | 15.2 |
| 3.26 | 53.7 | 14.0 | 16.0 |

[a] Col. (1) – d=3.2 Mpc.

[b] Col. (2) – Using rotation curve by Shostak (1973).

[c] Col. (3) – Using form by Kennicutt (1989, 1990).

[d] Col. (4) – $\Sigma_g = 1.45\,(\Sigma_{H_2} + \Sigma_{HI})$.

Table 4: Observed and Critical Gas Surface Densities of Individual Regions

| $\Delta\alpha(')^a$ | $\Delta\delta(')^a$ | R (kpc)$^b$ | $\Sigma_{H_2}$ (M$_\odot$ pc$^{-2}$) | $\Sigma_{HI}$ (M$_\odot$ pc$^{-2}$) | $\Sigma_g$ (M$_\odot$ pc$^{-2}$)$^c$ | $\Sigma_c$ (M$_\odot$ pc$^{-2}$)$^d$ |
|---|---|---|---|---|---|---|
| 2.0 | -0.5 | 2.3 | 3.7 | 8.2 | 17.3 | 18.0 |
| 1.5 | 0.0 | 2.0 | 5.1 | 8.8 | 20.2 | 19.6 |
| 1.5 | -0.5 | 1.6 | 4.3 | 8.0 | 17.8 | 22.2 |
| 1.5 | -1.0 | 1.7 | 4.0 | 7.8 | 17.1 | 21.5 |
| 1.0 | 1.0 | 2.6 | 5.3 | 7.8 | 19.0 | 16.7 |
| 1.0 | 0.0 | 1.3 | 6.0 | 8.0 | 20.3 | 24.5 |
| 1.0 | -0.5 | 1.1 | 5.5 | 7.2 | 18.4 | 26.2 |
| 0.5 | 0.0 | 0.7 | 4.8 | 7.0 | 17.1 | 30.3 |
| 0.5 | -0.5 | 0.7 | 6.9 | 6.9 | 20.0 | 30.3 |
| 0.0 | -0.5 | 0.8 | 6.8 | 7.1 | 20.0 | 29.2 |
| 0.0 | -1.0 | 1.6 | 6.3 | 7.1 | 19.4 | 22.2 |
| -1.0 | 1.0 | 1.4 | 3.6 | 6.2 | 14.2 | 23.7 |
| -1.0 | 0.5 | 1.1 | 8.7 | 6.6 | 22.2 | 26.2 |
| -1.5 | 1.0 | 1.7 | 3.6 | 6.3 | 14.4 | 21.5 |
| -3.0 | 1.5 | 3.2 | 4.2 | 8.2 | 18.0 | 14.2 |

$^a$ Cols. (1,2) – Right ascension and declination offsets are relative to $07^h\ 32^m\ 01.2^s$, $65°\ 42'\ 45''$ (1950).

$^b$ Col. (3) – d=3.2 Mpc.

$^c$ Col. (6) – $1.45(\Sigma_{H_2}+\Sigma_{HI})$.

$^d$ Col. (7) – Using form by Kennicutt (1989, 1990).

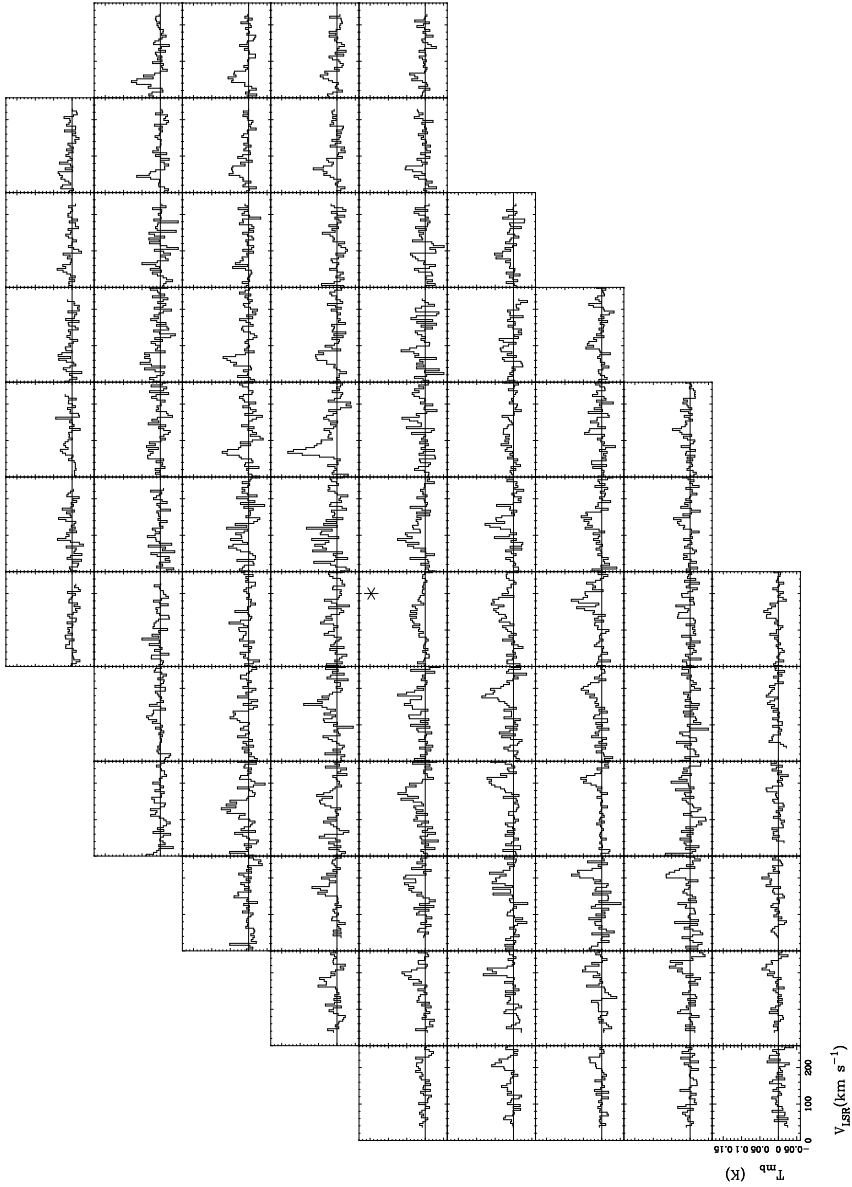

Fig. 1.— Grid of spectra covering the inner 3.2 kpc of NGC 2403. Brightness temperature is in units of $T_{mb}$. The asterisk marks the center position, $\alpha=07^h\ 32^m\ 01.2^s$, $\delta=65°\ 42'\ 45''$ (1950). North is upward and east is to the left. The observed positions are separated by $30''$ in both right ascension and declination.

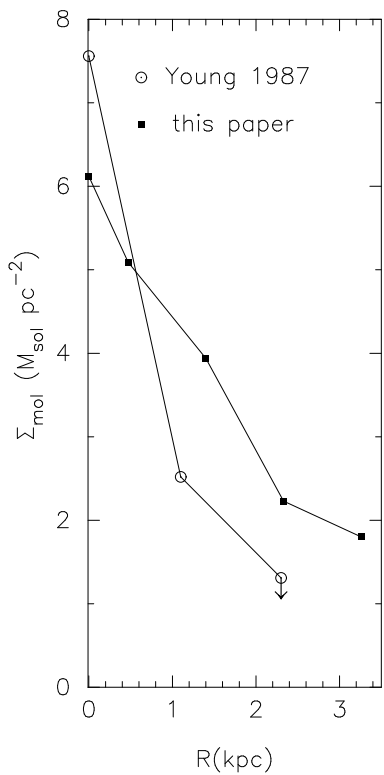

Fig. 2.— Radial profiles of $H_2$ surface densities. The squares trace the profile from this paper, and the open circles represent the profile taken from the major axis observations of Young (1987). Both profiles have been corrected for the inclination of the galaxy and assume a CO-to-$H_2$ conversion factor of $3\times 10^{20}$ cm$^{-2}$(K km s$^{-1}$)$^{-1}$.

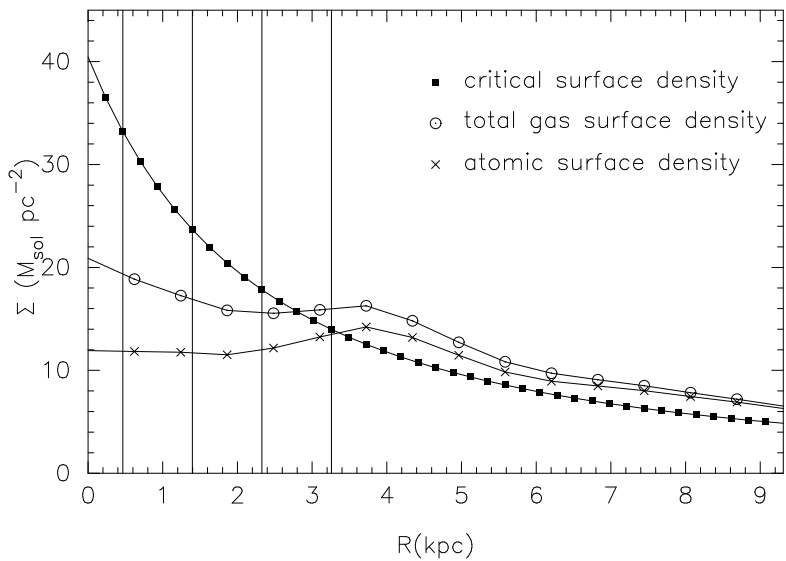

Fig. 3.— Radial profiles of HI, total gas, and critical surface densities in NGC 2403. The HI and total observed gas surface densities have been multiplied by 1.45 to account for other elements. The vertical lines mark the center of the annuli used to create a radial profile of the CO surface density from the observed fully sampled map.

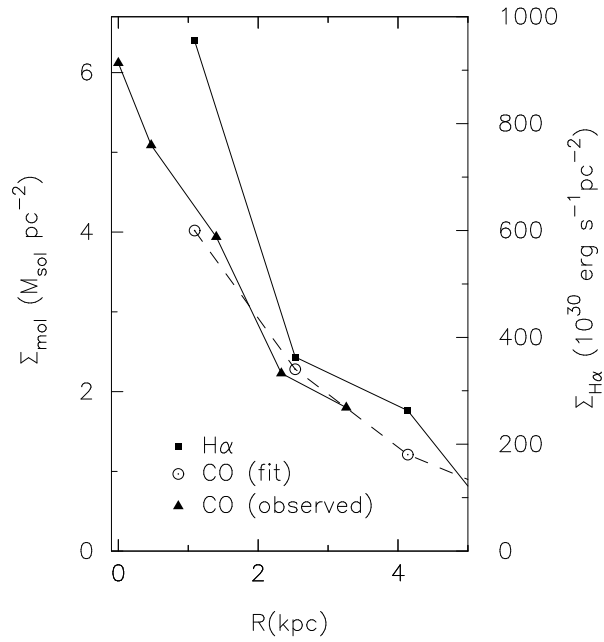

Fig. 4.— Radial profiles of Hα and CO surface densities. The triangles represent the observed CO emission, and the open circles connected by a dotted line represent the exponential fit to the observed CO data. The molecular surface density scale is shown along the left axis. The squares represent the Hα emission, with the Hα surface brightness scale shown along the right axis. See § 4.2 for details.